\documentclass[12pt]{iopart}
\pdfoutput=1
\usepackage[utf8]{inputenc} 			
\usepackage[T1]{fontenc}				
\usepackage[pdftex]{graphicx} 		    
\usepackage[svgnames]{xcolor}
\usepackage{cite}
\usepackage{iopams}                     
\usepackage{setstack}                   
\usepackage[english]{babel}
\usepackage{lmodern}
\usepackage{stmaryrd}
\usepackage{amssymb}
\usepackage{bbm}
\usepackage{tikz}
\usepackage{multirow}
\usepackage{yfonts}
\usepackage{subcaption}
\usepackage{float} 

\newcommand{\wh}{\widehat} 
\newcommand{\wtU}{\wh{U}}

\def\ee{e}
    
\def\to{\rightarrow}

\begin{document}

\title[Transitions driven by multibody interactions in an effective model of active matter]{Transitions driven by multibody interactions in an effective model of active matter}

\author{Thibaut Arnoulx de Pirey}

\address{Institut de Physique Théorique, CEA Saclay, Gif-sur-Yvette, 91190, France}

\author{Frédéric van Wijland}

\address{Laboratoire Matière et Systèmes Complexes,
              Université Paris Cité \& CNRS (UMR 7057),
              10 rue Alice Domon et Léonie Duquet,
              75013 Paris,
              France} 

\address{Yukawa Institute for Theoretical Physics, Kyoto University, Kitashirakawaoiwake-cho, Sakyo-ku, Kyoto, 606-8502, Japan}

\ead{thibaut.arnoulxdepirey@ipht.fr}
\date{\today\ -- \jobname}

\begin{abstract}
When out-of-equilibrium particles interact by means of pairwise forces, their stationary distribution in general exhibits many-body interactions. In the particular case of active particles, it has been shown numerically that the Motility Induced Phase Separation cannot be explained by the effective attraction emerging from two isolated particles, thereby highlighting the role of multibody interactions. In this work, we study the thermodynamics of the Fox-UCNA approximation for active particles interacting by means of pairwise repulsive forces. Working at large space dimension we establish that multibody interactions up to infinite order are instrumental in giving rise to such collective phenomena as phase transitions. We recover a MIPS-like first order transition, but also find a liquid-liquid transition at somewhat lower persistence times. This new transition is connected to a spin glass phase of orientational-like degrees of freedom with disordered interactions set by the particle positions themselves. 

\end{abstract}

\section{Introduction}
In the realm of Active Matter, the Motility Induced Phase Separation (MIPS)~\cite{cates2015motility} stands out as perhaps the simplest emergent collective phenomenon resting on the nonequilibrium features of the individual self-propelled particles that are involved. Such particles, interacting by means of isotropic and steeply repulsive forces, experience a phase separation between a dense and a dilute fluid phase at high enough density and persistence time. MIPS was first conjectured in 2008~\cite{tailleur2008statistical} based on a mesoscopic formulation in terms of a coarse-grained density field. It has since been observed in experimental systems, made of individual living constituents~\cite{self_p_bacteria} or of synthetic ones~\cite{PhysRevLett.110.238301}. In two~\cite{PhysRevLett.121.098003} and three~\cite{omar2021phase} space dimensions, the phase diagram  has been obtained with remarkable accuracy by means of numerical simulations, with a particular emphasis on the interplay between MIPS and the extension of the equilibrium crystallization to the nonequilibrium regime. Interestingly, recent studies on three dimensional systems~\cite{omar2021phase, turci2021phase} suggest that in  most of the phase space region where it is observed, the MIPS phase coexistence could actually be metastable whereas the true stationary state would be given by the coexistence of a disordered dilute phase and a dense, near-closed packed, crystalline one.

The similarities between the MIPS and standard liquid-gas phase separation of simple fluids in equilibrium interacting by means of a Lennard-Jones-like pair potential has been discussed in the literature. In particular, it has been argued by field theoretic treatments, and consistently with numerical simulations~\cite{turci2021phase, omar2021phase, maggi2021universality}, that the critical point of MIPS belongs to the Ising universality class~\cite{partridge2019}. From the liquid state theory viewpoint, however, there exists a fundamental difference, which is the subject of the present work, between self-propelled particle systems and simple fluids in equilibrium. In the latter, the $N$-body stationary equilibrium distribution is given by the Boltzmann weight and can be factorized over all the pairs of particles in the system,
\begin{eqnarray}
\mathcal{P}_N^{({\rm eq})}(\bi{r}_1, \dots , \bi{r}_N) \propto \prod_{i < j} \exp\left[-\beta U(\bi{r}_i-\bi{r}_j)\right] \, ,
\end{eqnarray}
with $\beta$ the inverse temperature and $U$ the pair potential. By contrast, for self-propelled particles in a large box of volume $V$, it is of course possible to write
\begin{eqnarray}
\ln\mathcal{P}_N^{({\rm eq})}(\bi{r}_1, \dots , \bi{r}_N)=\ln Z-\mathcal{H}_N(\bi{r}_1, \dots , \bi{r}_N)
\end{eqnarray}
where $Z$ is a constant such that when all particles are infinitely far away from each other (and from the walls of the box), $\mathcal{H}_N$ vanishes. The function $\mathcal{H}_N$ does not in general reduce~\cite{fodorhowfar2016} to a summation over distinct pairs of particles. It is thus convenient to split $\mathcal{H}_N$, for arbitrary $N$, into a sum of $k$-body interactions,
\begin{eqnarray}
\mathcal{H}_N(\bi{r}_1, \dots , \bi{r}_N)=\sum_{k=2}^N\sum_{i_1<\ldots<i_k} W^{(k)}(\bi{r}_{i_1},\ldots,\bi{r}_{i_k}) \,.
\end{eqnarray}
By requiring that the $W^{(k)}$'s are independent of $N$ the above equation defines these functions recursively without ambiguity according to 
\begin{eqnarray}\label{eq:defWIntro}
\fl
W^{(k+1)}(\bi{r}_{1},\ldots,\bi{r}_{k+1})=\mathcal{H}_{k+1}(\bi{r}_{1},\ldots,\bi{r}_{k+1})-\sum_{\ell=1}^k \sum_{i_1<\ldots<i_\ell}W^{(\ell)}(\bi{r}_{i_1},\ldots,\bi{r}_{i_\ell})\,.
\end{eqnarray}
In particular, this defines the effective pair interaction $W_{\rm eff}=W^{(2)}$ which we will discuss at length in this work. Our question of interest will be: what do we learn, if anything, from $W_{\rm eff}$?\\

A physical picture of the mechanism at work in MIPS was provided as soon as the phenomenon was predicted in \cite{tailleur2008statistical}: motility is decreased in regions of higher density, which feeds an instability mechanism leading to an accumulation of particles. However establishing the correct mesoscopic description in terms of a local density field based on the microscopics is a difficult endeavor that is usually bypassed by resorting to phenomenological arguments. To gain a complementary microscopic intuition of how MIPS unfolds,  it may be tempting, for self-propelled particles, to rely on an equilibrium-inspired understanding of the phase diagram, that of the liquid-gas phase separation of simple fluids in equilibrium. In such a picture, the role of $W_{\rm eff}$ is {\it a priori}  critical. A signature of self-propelled particles is their attraction to repulsive obstacles~\cite{ezhilan_alonso-matilla_saintillan_2015,wagner2017steady,lee2013active,elgeti2015run
}, and their self-attraction in the presence of pairwise repulsive forces. There are exact calculations of $W_{\rm eff}$ in one space dimension~\cite{slowman2016jamming}, and in higher dimensions at very high persistence~\cite{arnoulx2021active, de2023run} for hard particles, which show activity-induced attraction, consistently with other approximations~\cite{farage2015effective, marconi2015towards, marconi2016effective}. Indirect information can also be retrieved from the analysis of numerical simulations~\cite{ginot2015nonequilibrium}, where it was shown that, at low density, the pressure can be accurately described by the same equation as that of an equilibrium Baxter fluid with an activity-dependent adhesion strength. In most of these works the attractive part of $W_{\rm eff}$ was shown to be an increasing function of the persistence time. These results resonate with the idea that an attractive $W_{\rm eff}$ may be sufficient to account for the existence of phase separation. And indeed, on the analytical side, it is known that in one space dimension the effective attraction increases indefinitely as the persistence time is increased, similarly to the deepening of the attractive well in equilibrium (in any dimension) as the temperature is decreased.\\

It has however been shown in \cite{turci2021phase} that for two and three dimensional active Brownian particles, a purely pairwise approximation of the steady state distribution in the form
\begin{eqnarray}\label{eq:paireq_intro}
\mathcal{P}_N(\bi{r}_1, \dots , \bi{r}_N) \propto \prod_{i < j} \exp\left[- W_{\rm eff}(\bi{r}_i-\bi{r}_j)\right]
\end{eqnarray}
with $W_{\rm eff}$ defined through Eq.~\eref{eq:defWIntro}, does not induce attractive-enough interactions to account for MIPS at any persistence time. Multibody interactions are thus crucial in understanding the phase behavior of self-propelled particle systems. Recent analytical results in space dimensions $d\geq 2$~\cite{de2023run} corroborate this picture: in the dilute limit, the depth of the attractive well saturates at large persistence. The physical picture is that two active particles in contact will eventually skid past each other in finite time, irrespective of how large the persistence time is.\\

It is important to realize that multibody interactions control the phase diagram via an equation of state. To see this clearly, let us consider $N$ self-propelled particles confined by hard walls in a box of volume $V$ in $d$ space dimensions. We assume that the many-body probability distribution has the following form,
\begin{eqnarray}
\fl \mathcal{P}_N(\bi{r}_1, \dots , \bi{r}_N) = \frac{1}{Z} \frac{\exp\left[-\mathcal{H}_N(\bi{r}_1, \dots , \bi{r}_N)\right]}{N!}\prod_{i = 1}^N \prod_{\mu = 1}^d \Theta\left(r_i^\mu + \frac{L}{2}\right)\left(\frac{L}{2} - r_i^\mu\right) \, ,
\end{eqnarray}
with $V = L^d$ and where $\mathcal{H}_N$ is translationally and rotationally invariant. In other words, we assume that, when studying the bulk properties of the system, the effect of the confining walls can be represented by the constraints $-L/2 < r_i^\mu < L/2$ for all $i$ and $\mu$. We expect this assumption to be correct for torqueless spherically symmetric particles whose mechanical pressure follows an equation of state~\cite{PhysRevLett.114.198301}, and in the absence of boundary disorder~\cite{PhysRevE.105.044603}. The normalization constant is given by
\begin{eqnarray}
Z = \frac{1}{N!}\int_V \prod_{i = 1}^N \rm{d}^d \bi{r}_i  \exp\left[-\mathcal{H}_N(\bi{r}_1, \dots , \bi{r}_N)\right] \, ,
\end{eqnarray}
from which one can suggestively define the function
\begin{eqnarray}
F(N,V) = - \ln Z(N,V) \, ,
\end{eqnarray}
and an effective thermodynamic pressure
\begin{eqnarray}
P = - \left. \frac{\partial F}{\partial V} \right)_N \, .
\end{eqnarray}
In the absence of any kind of long-range interactions, which is bolstered by the fact that the critical point of MIPS belongs to the Ising universality class~\cite{partridge2019}, and for which the equivalence between the canonical and the grand canonical ensembles holds~\cite{ruelle1999statistical}, the phase behavior of the system can be inferred from a double tangent construction on $F$ or, equivalently, from a Maxwell construction on the effective thermodynamic pressure. In terms of $\mathcal{H}_N$, the latter can be obtained as~\cite{hill_1956}
\begin{eqnarray}
P = \rho \left( 1 - \frac{1}{d} \left\langle \frac{1}{N}\sum_i \bi{r}_i \cdot \bnabla_{\bi{r}_i} \mathcal{H}_N \right\rangle \right) \, .
\end{eqnarray}
Therefore, unlike the standard expectation for an equilibrium system with pairwise interactions, we see that the spatial pair distribution function alone does not control the phase behavior of self-propelled particles. We will see this construction at play in a context where this multibody feature is preserved while allowing for analytical progress. \\

In this work, we investigate the importance of multibody interactions in the steady state of an approximate model for active matter, the Unified Colored Noise Approximation (UCNA)~\cite{PhysRevA.35.4464} of the Active Ornstein-Ulhenbeck (AOU) dynamics. This approximation, which shares the same equilibrium distribution~\cite{wittmann2017effective} (in the absence of an additional thermal noise) as the Fox one~\cite{fox1986functional}, is one of the main approximations of active systems interacting via pairwise forces~\cite{farage2015effective, maggi2015multidimensional, marconi2016effective, wittmann2017effective} as the steady-state distribution can be derived explicitly at any persistence time (and it becomes exact in the small persistence-time limit~\cite{fodorhowfar2016}). Interestingly, this many-body stationary distribution displays multibody interactions. 
 
It however remains in general a difficult task to derive, from the many-body distribution, the thermodynamic properties of the system. A mean-field study of the UCNA stationary distribution has been proposed in \cite{paoluzzi2020statistical}, but relied on a technical assumption (namely the substitution $\ln\det\left(\mathbbm{1} + H\right) \to {\rm Tr} H$ where the matrix $H$ will be introduced in Eq.~\eref{eq:UCNAstat}) which misses part of the physics by making the emergence of effective attraction a two-body rather than a collective feature. Here we follow a different path, and obtain the phase diagram of the model in the limit of large space dimension, known to allow for significant analytical progress in the study of interacting particle systems. In equilibrium, the equation of state of hard spheres has been derived~\cite{frisch1985classical} in this limit, and phase transitions in more complex fluids~\cite{carmesin1989liquid,carmesin1991binary} have been described exactly. The limit of infinite dimension has also been instrumental in the study of the glass and jamming transitions, both from static~\cite{parisi2010mean} and dynamical~\cite{maimbourg2016solution} points of view. All these approaches however heavily rely on the factorization property of the Boltzmann measure or of the dynamical partition function (in the Janssen-de Dominicis formalism), which leads to an exact truncation to second order (up to exponentially small corrections in the dimension) of the virial expansion of the free energy functional. In this work, we proceed by expressing the steady-state measure in the UCNA as a product over pair functions in an extended phase space with auxiliary variables, thus allowing us to resort to the machinery of the Mayer expansion and of its truncation in $d \to \infty$. As a result, we will see that the density expansion of the free energy in the UCNA retains terms up to infinite order.\\

We begin by reviewing the UCNA scheme as applied to a model of interacting self-propelled particles and explain how the infinite-dimensional limit is to be considered. We then show that the two-body physics cannot account on its own for the existence of a phase transition. Then we determine the free energy and use it to extract the phase diagram in the density-persistence time variables. Our main results are that there exists a first-order MIPS-like phase separation at large persistence times, and that at moderate persistence times, a continuous liquid-liquid phase transition occurs.

\section{The Unified Colored Noise Approximation}
A standard model of self-propelled particles in the active matter literature is the active Ornstein-Ulhenbeck one (see  \cite{szamel2014self, martin2021statistical} and references therein). For $N$ interacting particles in dimension $d$ labeled by $i \in \llbracket 1, N \rrbracket$, the dynamics of the $i^{\, th}$ particle follows
\begin{eqnarray} \label{eq:AOUP1}
    \dot{\bi{r}}_i  = \bi{v}_i - \sum_{j \neq i} \bnabla_i U(\bi{r}_i - \bi{r}_j) \, ,
\end{eqnarray}
where $U$ is the pairwise interparticle potential and the active driving $\bi{v}_i$ is modeled as a $d$-dimensional Ornstein-Ulhenbeck process
\begin{eqnarray} \label{eq:AOUP2}
    \dot{\bi{v}}_i = - \frac{\bi{v}_i}{\tau} + \frac{\sqrt{2D}}{\tau}\bfeta_i(t) \, ,
\end{eqnarray}
with $\bfeta_i(t)$ a Gaussian white noise with correlations $\left\langle \eta_i^\mu(s) \eta_j^\nu(t) \right\rangle = \delta_{ij}\delta^{\mu\nu}\delta(t - s)$. In the steady state, the process $\bi{v}_i(t)$ is exponentially correlated in time with correlation time $\tau$. The pair potential $U$ is assumed to be radially symmetric, $U(\bi{r} ) = U(r)$ with $r = \vert  \bi{r} \vert$. Introducing the velocity $\bi{p}_i = \dot{\bi{r}}_i$, the dynamics \eref{eq:AOUP1}-\eref{eq:AOUP2} can be written as
\begin{eqnarray} \label{eq:AOUP3}\fl
\tau \dot{p}_i^\mu + \left[\delta^{\mu\nu} + \tau \sum_{j \neq i}\partial^\mu\partial^\nu U(\bi{r}_i - \bi{r}_j) \right]p_i^\nu = - \sum_{j \neq i}\partial^\mu U(\bi{r}_i - \bi{r}_j) + \sqrt{2D} \, \eta_i^\mu(t) \, ,
\end{eqnarray}
where repeated indices are summed over. Equation~\eref{eq:AOUP3} is at the basis of the UCNA of the active Ornstein-Ulhenbeck dynamics. The approximation consists concretely in dropping the inertial term $\tau \dot{p}_i^\mu$ to arrive at the overdamped equation of motion
\begin{eqnarray}\label{eq:UNCAdyn}
     \left[\delta^{\mu\nu} + \tau \sum_{j \neq i}\partial^\mu\partial^\nu U(\bi{r}_i - \bi{r}_j) \right]\dot{r}_i^\nu = - \sum_{j \neq i}\partial^\mu U(\bi{r}_i - \bi{r}_j) + \sqrt{2D} \, \eta_i^\mu(t) \, ,
\end{eqnarray}
understood as being Stratonovich-discretized. The Stratonovich prescription allows the dynamics in Eq.~\eref{eq:UNCAdyn} to be consistent with the original dynamics Eq.~\eref{eq:AOUP1}-\eref{eq:AOUP2} to first order in $\tau$ at small persistence time. The resulting dynamics Eq.~\eref{eq:UNCAdyn} is an equilibrium dynamics which renders possible the computation of the stationary distribution with respect to which detailed balance holds~\cite{maggi2015multidimensional}:
\begin{eqnarray} \label{eq:UCNAstat}
    P_s(\{\bi{r}_i\}) = \frac{1}{Z}\, \rme^{- \beta \sum_{i < j}U(\bi{r}_i-\bi{r}_j) - \frac{\beta \tau}{2}\sum_i \left(\sum_{j\neq i}\bnabla_{i}U(\bi{r}_i-\bi{r}_j)\right)^2}\left|\det\left(\mathbbm{1} + H\right)\right| \, ,
\end{eqnarray}
with $Z$ the normalization constant, $\beta = D^{-1}$ the actual temperature of the zero persistence time dynamics and where $H$ is an $N d \times N d$ matrix with coefficients
\begin{eqnarray} \label{eq:matrixH}
H_{i\alpha, j \beta} = \tau \left( \sum_{k \neq i}\partial^\alpha \partial^\beta U(\bi{r}_i - \bi{r}_k) \delta_{ij} - \partial^\alpha \partial^\beta U(\bi{r}_i - \bi{r}_j) \left[1 - \delta_{ij}\right]\right) \,.
\end{eqnarray}
The probability distribution in Eq.~\eref{eq:UCNAstat} agrees to first order in a small $\tau$ expansion with the stationary distribution of the original active Ornstein-Ulhenbeck dynamics~\cite{fodorhowfar2016}. Equation~\eref{eq:UCNAstat} can nevertheless be approached non-perturbatively in $\tau$. It is at the basis of our analytical study on the importance of multibody interactions in the phase diagram of active matter systems.

\subsection{The infinite dimensional limit}\label{subsec:ucna_scalings}
Following the pioneering approach of \cite{frisch1985classical} for the liquid phase of classical hard spheres, and later successfully extended to the statics and dynamics of glassy systems~\cite{kurchan2012exact, maimbourg2016solution}, we study analytically the thermodynamic properties of the position space stationary measure in Eq.~\eref{eq:UCNAstat} in the limit of infinite space dimension $d \to \infty$. The scaling of the model parameters with the dimension follow from \cite{agoritsas2019out}. The interaction potential is assumed to be short-ranged and steeply repulsive for $r < \sigma$ and scales as
\begin{eqnarray}\label{eq:potential_scaling}
        U(r) = \wtU(h) \, \mbox{  with  } \, h = d\left(r/\sigma - 1\right) \, ,
\end{eqnarray}
where $\sigma$ is the effective diameter of a particle. The coefficient $\beta$ is kept fixed so as to maintain $\beta U(r)$ finite. Furthermore, each particle is assumed to have roughly $d$ neighbors, \textit{i.e.} for each particle there are $O(d)$ particles with which the rescaled interaction potential $\wtU$ is non-vanishing. This requires the density $\rho$ to scale as
\begin{eqnarray}\label{eq:density_scaling}
    \frac{\rho \mathcal{V}_d(\sigma)}{d} = \wh{\varphi} \, ,
\end{eqnarray}
with $\wh{\varphi}$ finite and where $\mathcal{V}_d(\sigma)$ is the volume of the $d$-dimensional ball of radius $\sigma$. Lastly, the correlation time $\tau$ is scaled as $\tau = \hat{\tau}/d^2$. This guarantees that, over timescales of order $\tau$, the variations of $r_i^\mu$ scale as $O(1/d)$, which is the natural length scale in light of Eq.~\eref{eq:potential_scaling}.

\section{Absence of transition at the two body level} \label{sec:notransiUCNA} 
\begin{figure}[!h]
        \centering
        \includegraphics[width=0.47\textwidth]{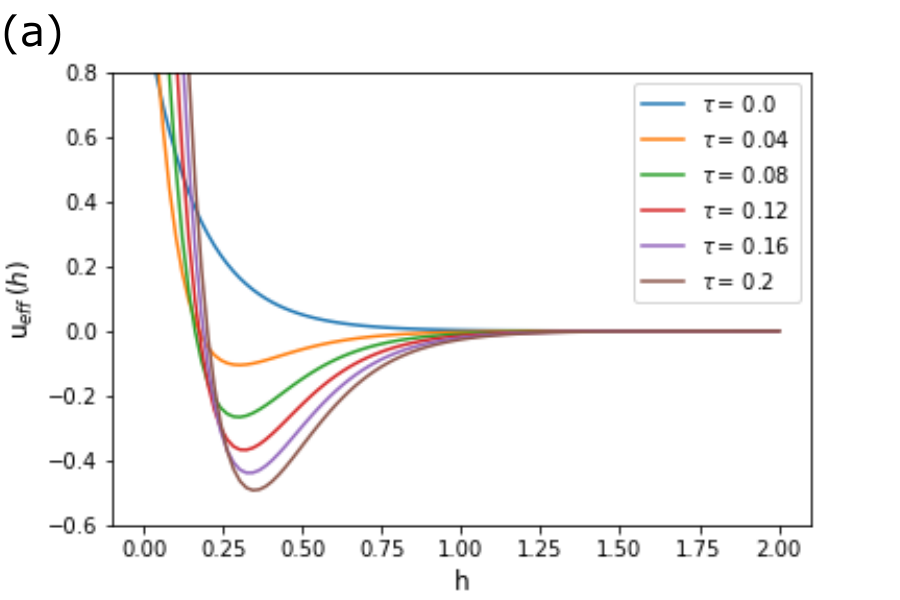}
        ~
        \includegraphics[width=0.47\textwidth]{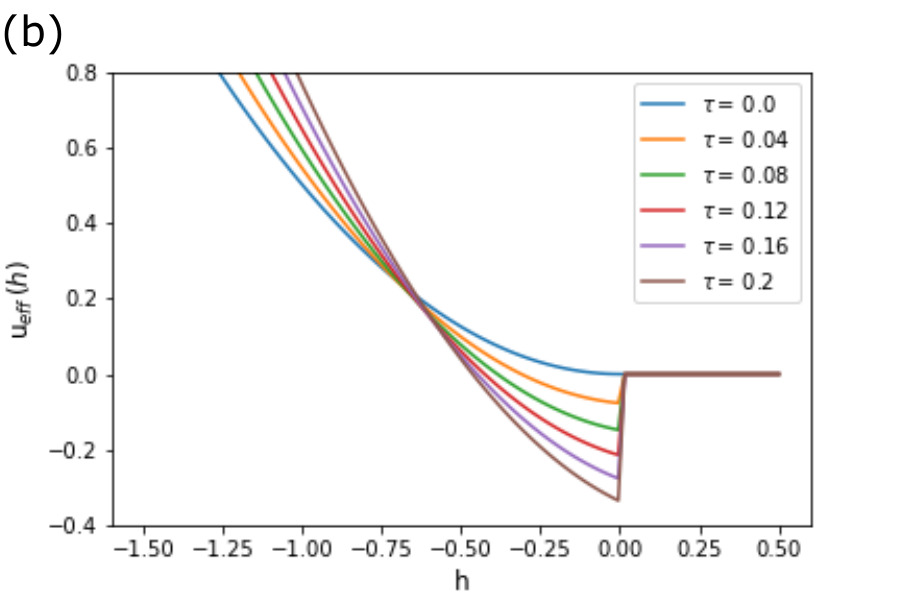}
        \caption{Effective pair interaction for different values of $\hat{\tau}$ for $\beta = 1$ and $\sigma = 1$. (a) The pair potential is taken of the form $\wtU(h) = u_0 \, \ee^{- \lambda h}$ with $u_0 = 1$ and $\lambda = 6$. (b) The pair potential is taken of the form $\wtU(h) = u_0 \, h^2 \Theta(-h)/2$ with $u_0 = 1$.}
        \label{fig:eff_potentials}
    \end{figure}

\begin{figure}[!h]
        \centering
        \includegraphics[width=0.47\textwidth]{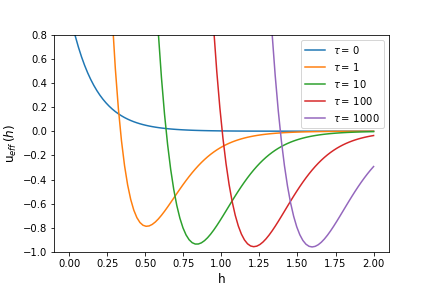}
        \caption{Effective pair potential for different values of $\hat{\tau}$ for $\beta = 1$ and $\sigma = 1$ and $\wtU(h) = u_0 \, \ee^{- \lambda h}$ with $u_0 = 1$ and $\lambda = 6$. The depth of the attractive well saturates as $\hat{\tau}$ increases.}
        \label{fig:effpair_satur}
\end{figure}

Before studying the full phase diagram of Eq.~\eref{eq:UCNAstat}, we restrict, in the present section, our attention to the case $N = 2$. This allows us to define the effective pair potential $W_{\rm eff}(\bi{r})$ as
\begin{eqnarray}
    P_{s}^{N = 2}(\bi{r}_1, \bi{r}_2) \propto \ee^{-W_{\rm eff}(\bi{r}_1 - \bi{r}_2)} \, ,
\end{eqnarray}
defined in such a way that $\lim_{|\bi{r}| \to \infty}W_{\rm eff}(\bi{r}) = 0$. $W_{\rm eff}(\bi{r})$ also describes the structure of a thermodynamic system in the limit of vanishing density $\rho$. It can be inferred from Eq.~\eref{eq:UCNAstat} and was obtained in \cite{maggi2015multidimensional} as
\begin{eqnarray} \label{eq:effective_pair}
\fl
W_{\rm eff}(r) =  \beta U(r) + \beta \tau U'(r)^2 - (d - 1)\ln \left|1 + 2 \tau \frac{U'(r)}{r}\right| - \ln\left|1 + 2 \tau U''(r)\right| \, .
\end{eqnarray}
In a small $\tau$ expansion, the above expression leads to
\begin{eqnarray} \label{eq:effective_pair_small_tau}
W_{\rm eff}(r) = U(r) + \tau U'(r)^2 - 2 \tau (d-1) \frac{U'(r)}{r} - 2 \tau U''(r) + O(\tau^2) \, .
\end{eqnarray}
Let us assume that the interaction potential $U(r)$ is purely repulsive and convex. The first three terms in the right-hand side of Eq.~\eref{eq:effective_pair_small_tau} then correspond to purely repulsive contributions. The last one, however, corresponds to activity-induced attraction. At arbitrary $\tau$, and in dimension $d > 1$, the formula in Eq.~\eref{eq:effective_pair} however might yield unphysical results for broad classes of potentials~\cite{wittmann2017effective} due to the appearance of negative eigenvalues in the spectrum of $\mathbbm{1} + H$. This has led various authors~\cite{farage2015effective,marconi2016effective,wittmann2017effective} to perform additional approximations to construct an effective pair potential and to study the resulting collective properties in a pairwise approximation {\it à la} Eq.~\eref{eq:paireq_intro}. This procedure was however shown to overestimate the effective attraction and to have a limited range of quantitative and qualitative validity~\cite{rein2016applicability}. This problem of unphysical effective pair potential is cured in the limit of infinite dimension in which the effective pair interaction becomes well behaved for convex potentials at arbitrary values of $\hat{\tau}$ and $\beta$ as we obtain
\begin{eqnarray}
\label{eq:effpairpot}
    \widehat{W}_{\rm eff}(h) = \beta \wtU(h) + \frac{\beta \hat{\tau}}{\sigma^2}\wtU'(h)^2 -  \frac{2\hat{\tau}}{\sigma^2}\wtU'(h) - \ln \left\vert 1 + \frac{2\hat{\tau}}{\sigma^2} \wtU''(h) \right\vert \, .
\end{eqnarray}
In \fref{fig:eff_potentials}, we plot the effective pair interaction for different values of $\hat{\tau}$ in the case of an exponentially repulsive pair potential $\wtU(h) = u_0 \, \ee^{- \lambda h}$ (left panel) and of a harmonic sphere one $\wtU(h) = u_0 \, h^2 \Theta(-h)/2$ (right panel). At $\hat{\tau} = 0$ the effective pair interaction equals $\beta$ times the pair potential. For large enough $\hat{\tau} >0$, the effective interaction develops an attractive part due to the logarithmic term in Eq.~\eref{eq:effpairpot}.
Interestingly, as we shall now see, it turns out that this effective pair interaction, while displaying, for convex potentials, an attractive part due to the logarithmic term, is not able on its own to induce phase separation. Let us consider a system of $N$ particles with position space stationary distribution retaining only the pair interactions $W_{\rm eff}$,
\begin{eqnarray}\label{eq:stationary2Body}
\mathcal{P}\left(\bi{r}_1, \dots, \bi{r}_N\right) = \frac{1}{Z} \prod_{i < j} \rme^{-W_{\rm eff}\left(\bi{r}_i - \bi{r}_j\right)} \, ,
\end{eqnarray}
In the limit of infinite space dimension, the free energy functional is truncated to second order in a density expansion~\cite{frisch1985classical, kurchan2012exact}
\begin{eqnarray}
        \mathcal{F}[\rho(\bi{r})] = \int \rmd \bi{r} \, \rho(\bi{r}) \left( \ln \rho(\bi{r}) - 1 \right) - \frac{1}{2}\int \rmd \bi{r} \, \rmd \bi{r}' \, \rho(\bi{r}) \rho(\bi{r}') f\left(\bi{r}, \bi{r}'\right) \, ,
\end{eqnarray}
up to corrections that are exponentially small in $d$ and where $f(\bi{r}) = \ee^{-W_{\rm eff}(\bi{r})} - 1$ is the Mayer function. For a homogeneous phase with rescaled density $\wh{\varphi}$, the thermodynamic pressure is thus quadratic in the density and reads
\begin{eqnarray}
   \frac{P(\wh{\varphi})\mathcal{V}_d(\sigma)}{d} & = \wh{\varphi} \left( 1 + d \wh{\varphi} B(\hat{\beta},\hat{\tau})\right) \\ & \simeq  d \,  \hat{\varphi}^2 B(\beta,\hat{\tau}) \, ,
\end{eqnarray}
where $B$ is the second virial coefficient,
\begin{eqnarray}
    B(\beta,\hat{\tau}) = \frac{1}{2} \int \rmd h \, \ee^h \left( 1 - \rme^{-\,\wh{W}_{\rm eff}(h)}\right) \, .
\end{eqnarray}
Let us now investigate the case of harmonic spheres $\wtU(h) =(u_0/2) h^2 \Theta(-h)$ and introduce $c_1 = 2\hat{\tau} u_0/\sigma^2$ and $c_2 = \beta u_0$. The latter is the ratio between the pair potential energy scale and the effective temperature of the $\tau = 0$ dynamics whereas $\sqrt{c_1/c_2}$ controls the ratio between the run length of the original non-interacting active Ornstein-Ulhenbeck dynamics and the size of a particle. The effective potential then reads,
\begin{eqnarray}
    \wh{W}_{\rm eff}(h) = \left[ c_2(1+c_1)\frac{h^2}{2} - c_1 h - \ln\left(1+c_1\right) \right] \Theta(-h) \, ,
\end{eqnarray}
and the coefficient $B$ has the explicit expression
\begin{eqnarray}
\fl
    2 B(c_1,c_2) = 1 - \sqrt{\pi}\exp{\left(\frac{1+c_1}{2c_2}\right)}{\rm erfc}\left(\sqrt{\frac{1+c_1}{2c_2}}\right)\sqrt{\frac{1+c_1}{2 c_2}} > 0 \qquad \forall \, c_1, c_2 > 0 \, .
\end{eqnarray}
Thus, $B(c_1,c_2)$ being always positive, homogeneous phases are stable in the whole parameter space. Similar conclusions can be drawn for the potential used to construct the left panel of \fref{fig:eff_potentials}. Effective pair interactions alone are thus not able to account for phase separation. This may come as a surprise given the shape of the effective pair potential (especially the one on the left panel of \fref{fig:eff_potentials}) that visually resembles the standard Lennard-Jones potential that is well known to account for phase separation in equilibrium simple liquids. However, in such a Lennard-Jones system, the depth of the attractive well of the effective pair interaction can be made arbitrarily large by lowering the temperature. This is not the case here as the depth of the attractive part of the effective pair interaction saturates at large $\tau$, see \fref{fig:effpair_satur}). This is reminiscent of the results obtained in \cite{arnoulx2021active} where it was shown that the dilute limit of the pair distribution function of self-propelled hard spheres has a well-defined limit in the distribution sense in the limit of large persistence time. We believe that this observation is key to understanding the numerical results presented in \cite{turci2021phase} about the role of multibody interactions in active particle systems, which we investigate next within the context Eq.~\eref{eq:UCNAstat}.

\section{Computing the free energy}\label{sec:computation}

This section is devoted to the computation of the free energy of the UCNA in the limit of infinite dimension. The physical implications of these results follow up in section \ref{sec:phase_diagram}. The present derivation can be read independently from the rest of the paper.

\subsection{Mapping towards a pairwise interacting system}\label{sec:mapping_pair}  

To extract the thermodynamic properties of the measure in Eq.~\eref{eq:UCNAstat}, we use the grand canonical ensemble and introduce the associated partition functional $\Xi[\mu]$ defined by
\begin{eqnarray} \label{eq:grand_canonical_UCNA}
\fl \Xi[\mu] = \sum_{N = 0}^{+\infty} \frac{1}{N!}\int \left[\prod_{i} \rmd \bi{r}_i \, \rme^{\, \mu(\bi{r}_i)}\right] \rme^{- \beta \sum\limits_{i < j}U(\bi{r}_i-\bi{r}_j) - \frac{\beta \tau}{2}\sum\limits_i \left(\sum\limits_{j\neq i}\bnabla_{i}U(\bi{r}_i-\bi{r}_j)\right)^2}  \left|\det\left(\mathbbm{1} + H\right)\right| \, ,
\end{eqnarray}
with $\mu(\bi{r})$ the generalized chemical potential. As we will see later, the density scaling in \eref{eq:density_scaling} imposes to keep $\hat{\mu} = (\mu - \ln(d/\mathcal{V}_d(\sigma))/d$ finite as $d\to\infty$. The expression in Eq.~\eref{eq:grand_canonical_UCNA} is not suitable for directly taking the infinite dimensional limit because of the presence of multibody interactions. We thus map the model under consideration onto one with purely pairwise interactions, which is possible on condition of working in an extended phase space. To proceed, we first note that $\Xi[\mu] = \lim_{n\to0} \Xi_n[\mu]$ where
\begin{eqnarray} \label{eq:grand_canonical_UCNA_n}
\fl \Xi_n[\mu] = \sum_{N = 0}^{+\infty} \frac{1}{N!}\int \left[\prod_{i} \rmd \bi{r}_i \, \rme^{\, \mu(\bi{r}_i)}\right] \rme^{- \beta \sum\limits_{i < j}U(\bi{r}_i-\bi{r}_j)} \left[\frac{\rme^{\frac{\beta \tau}{2}\sum\limits_i \left(\sum\limits_{j\neq i}\bnabla_{i}U(\bi{r}_i-\bi{r}_j)\right)^2}}{\left|\det\left(\mathbbm{1} + H\right)\right|}\right]^{n-1}  \, .
\end{eqnarray}
The three-body interaction term in the numerator is disentangled by introducing $N$ $d$-dimensional Gaussian vectors $\bpsi_i$ through
\begin{eqnarray}
\fl \exp\left(\frac{\beta \tau}{2}\sum\limits_i \left(\sum\limits_{j\neq i}\bnabla_{i}U(\bi{r}_i-\bi{r}_j)\right)^2 \right) = \int \left[\prod_i \frac{\rmd \bpsi_i \, \rme^{- \frac{\bpsi_i^2}{2}}}{\left(2 \pi\right)^{d/2}}\right]\rme^{- \sqrt{\beta \tau} \sum\limits_{i<j} \left(\bpsi_i - \bpsi_j\right)\cdot \bnabla_iU\left(\bi{r}_i - \bi{r}_j\right) } \, . 
\end{eqnarray} 
We proceed similarly to disentangle the all-to-all interaction term corresponding to the determinant. In the following, we assume that the matrix $\mathbbm{1} + H$ is positive definite. The analysis of the effective pair potential in section \ref{sec:notransiUCNA} indeed suggests that the set of points where this approximation breaks down has very small measure in the limit $d \to \infty$. We introduce a set of $N$ $d$-dimensional vectors $\boldsymbol{\phi}_{i}$ and use a Gaussian integral
\begin{eqnarray}
\sqrt{\frac{1}{\det\left(\mathbbm{1}+H\right)}}=\int\left[\prod_{i=1}^{N}\frac{{\rm d}\boldsymbol{\phi}_{i}\,{\rm e}^{-\frac{\boldsymbol{\phi}_{i}^{2}}{2}}}{\left(2\pi\right)^{d/2}}\right]\exp\left(-\frac{1}{2}\sum_{i,j=1}^{N}\phi_{i}^{\alpha}\phi_{j}^{\beta}H_{i\alpha,j\beta}\right) \,.
\end{eqnarray}
To keep notations concise, we use Greek indices and Einstein summation convention for the spatial dimensions. For  $n \in \mathbbm{N}$ with $n > 1$, the partition functional $\Xi_n[\mu]$ can thus be evaluated by introducing the replicated fields $\{\bpsi_i^a\}_{a = 1, \dots, n-1}$ and $\{\bphi_i^a\}_{a = 1, \dots, 2(n-1)}$. The resulting theory is that of a system with purely pairwise interactions whose partition functional reads
\begin{eqnarray} \label{eq:xi_n}
\fl \Xi_{n}[\mu]	=\sum_{N=0}^{+\infty}\frac{1}{N!}\int\left[\prod_{i}{\rm d}\boldsymbol{r}_{i}\prod_{a=1}^{n-1}\frac{{\rm d}\boldsymbol{\psi}_{i}^{a}}{(2\pi)^{d/2}}\prod_{a=1}^{2n-2}\frac{{\rm d}\boldsymbol{\phi}_{i}^{a}}{(2\pi)^{d/2}}\right]\prod_{i}\exp\left(\mu(\boldsymbol{r}_{i})-\sum_{a=1}^{n-1}\frac{\left(\boldsymbol{\psi}_{i}^{a}\right)^{2}}{2}-\sum_{a=1}^{2n-2}\frac{\left(\boldsymbol{\phi}_{i}^{a}\right)^{2}}{2}\right) \nonumber
	 \\ \fl \prod_{i<j}\exp\left(-\beta U(\boldsymbol{r}_{i}-\boldsymbol{r}_{j})-\sqrt{\beta\tau}\sum_{a=1}^{n-1}\left(\boldsymbol{\psi}_{i}^{a}-\boldsymbol{\psi}_{j}^{a}\right)\cdot\boldsymbol{\nabla}U(\boldsymbol{r}_{i}-\boldsymbol{r}_{j})\right. \nonumber \\ \left. -\frac{\tau}{2}\sum_{a=1}^{2n-2}\left(\phi_{i}^{a,\alpha}-\phi_{j}^{a,\alpha}\right)\left(\phi_{i}^{a,\beta}-\phi_{j}^{a,\beta}\right)\partial^{\alpha}\partial^{\beta}U(\boldsymbol{r}_{i}-\boldsymbol{r}_{j})\right) \,,
\end{eqnarray}
As we explain below, this free energy functional can be computed exactly in the $d \to \infty$ limit. The continuation to $n \to 0$ is taken at the end. The validity of the analytic continuation and the possibility to invert the $n \to 0$ and $d \to \infty$ limits  is not discussed in this work.

\subsection{The free energy functional}
It is useful to group the auxiliary degrees of freedom into a single field $\bxi_i = \{\bpsi_i^a, \bphi_i^a\}$ and define the measure
\begin{eqnarray}
{\rm d}\bxi_i = \prod_{a=1}^{n-1}{\rm d}\boldsymbol{\psi}_{i}^{a}\prod_{a=1}^{2n-2}{\rm d}\boldsymbol{\phi}_{i}^{a} \,.
\end{eqnarray}
To compute the free energy functional, we promote $\Xi_n[\mu]$ from a functional of the chemical potential $\mu(\bi{r})$ to a functional of the generalized chemical potential $j(\bi{r}, \bxi)$ as
\begin{eqnarray} \label{eq:xi_n_generalized}
\Xi_n[j] = \sum_{N = 0}^{+\infty} \frac{1}{N!} \! \int \prod_{i = 1}^N \rmd \bi{r}_i \, \rmd \bxi_i  \left[ \prod_{i = 1}^N \ee^{j(\bi{r}_i, \bxi_i)} \right]  \prod_{i < j} \left( 1 + f(\bi{r}_i, \bxi_i ; \bi{r}_j, \bxi_j) \right)  \, ,
\end{eqnarray}
where the Mayer function $f$ can be read from \eref{eq:xi_n}. We are interested in the properties of this model at the physical generalized chemical potential 
\begin{eqnarray}
j^{*}(\boldsymbol{r},\boldsymbol{\xi})=\mu-\sum_{a=1}^{n-1}\frac{\left(\boldsymbol{\psi}_{i}^{a}\right)^{2}}{2}-\sum_{a=1}^{2n-2}\frac{\left(\boldsymbol{\phi}_{i}^{a}\right)^{2}}{2}-\frac{3d(n-1)}{2}\ln(2\pi) \,.
\end{eqnarray}
Of particular interest is the one-body density $\rho^*(\bi{r},\bxi)$ given by
\begin{eqnarray}
\rho^*(\bi{r},\bxi) = \left.\frac{\delta \ln \Xi_n[j]}{\delta j(\bi{r},\bxi)}\right|_{j^*(\bi{r}, \bxi)} \,.
\end{eqnarray}
The latter can be obtained by minimizing the free energy functional $\mathcal{F}_n[\rho]$ defined from the Legendre transform of the logarithm of the grand canonical partition functional,
\begin{eqnarray}
\rho^*(\bi{r},\bxi) = \underset{\rho(\bi{r},\bxi)}{\rm argmin}\,\mathcal{F}_n[\rho]\,,
\end{eqnarray}
with 
\begin{eqnarray}
\mathcal{F}_n[\rho] = \underset{j(\bi{r},\bxi)}{\rm max} \left[-\ln \Xi_n[j] + \int \rmd \bi{r} \rmd \bxi \, \rho(\bi{r},\bxi)\left(j(\bi{r},\bxi) - j^*(\bi{r},\bxi)\right)\right] \,.
\end{eqnarray}
Remarkably, in the limit of infinite spatial dimension, the density expansion of the free energy functional is truncated to second order, so that 
\begin{eqnarray}\label{eq:free_energy}
    \mathcal{F}_n[\rho] = & \int \rmd \bi{r} \rmd \bxi \, \rho(\bi{r},\bxi)\left[\ln \rho(\bi{r},\bxi) - 1 - j^*(\bi{r},\bxi)\right] \nonumber \\ & -  \frac{1}{2} \int \rmd \bi{r} \, \rmd \bxi \, \rmd \bi{r}' \, \rmd \bxi' \, \rho(\bi{r},\bxi) \rho(\bi{r}',\bxi') f(\bi{r}, \bxi ; \bi{r}', \bxi') \, .
\end{eqnarray}
Therefore, the physical one-body distribution $\rho^*(\bi{r}, \bxi)$ is a solution of the stationarity equation
\begin{eqnarray} \label{eq:one_body}
  \rho^*(\bi{r}, \bxi) = \ee^{\,j^*(\bi{r},\bxi)}\exp{\left(\int \rmd\bi{r}' \rmd \bxi' \rho^*(\bi{r}', \bxi') f(\bi{r}, \bxi ; \bi{r}', \bxi') \right)}\,.
\end{eqnarray}
Fortunately, similarly to what happens in glasses~\cite{kurchan2012exact}, the details of the physical one-body distribution $\rho^*(\bi{r},\bxi)$ are not needed to predict the phase diagram of the system nor its structure factor. Indeed, to leading order in the dimension $d$, the free energy functional $\mathcal{F}_n[\rho]$ depends only on a finite number of moments of the distribution $\rho(\bi{r}, \bxi)$. The macroscopic properties of the system under study can therefore be inferred from a stationarity condition on this set of moments rather than dealing with the full distribution that generates them. This will be the subject of the rest of this section.

\subsection{Scaling of the one-body distribution}
We are interested in the description of homogeneous and rotationally invariant phases. We thus introduce the $3(n-1)\times3(n-1)$ matrix of scalar products $S$ 
\begin{eqnarray}
S=\left(\begin{array}{cc}
M & R\\
R^{T} & Q
\end{array}\right)
\end{eqnarray}
with $M_{ab}	=\boldsymbol{\psi}_{a}\cdot\boldsymbol{\psi}_{b}$, $Q_{ab}	=\boldsymbol{\phi}_{a}\cdot\boldsymbol{\phi}_{b}$ and $R_{ab}	=\boldsymbol{\psi}_{a}\cdot\boldsymbol{\phi}_{b}$. We further assume the existence of an underlying large deviation principle and optimize the free energy over density functions respecting the following ansatz
\begin{eqnarray}\label{eq:ansatz}
\rho(\boldsymbol{r},\boldsymbol{\xi})=\rho\exp\left(d\,\Gamma\left(\frac{S}{d}\right)\right) \,,
\end{eqnarray}
with $\Gamma$ an $O(1)$ $d$-independent scaling function. The validity of this ansatz is corroborated at the end by the consistency of the resulting optimization principle. Before going further, we note that the normalization condition 
\begin{eqnarray}
1=\int d\boldsymbol{\xi}\exp\left(d\,\Gamma\left(\frac{S}{d}\right)\right) \,,
\end{eqnarray}
becomes after performing the change of variables $\bxi \to \sqrt{d}\bxi$, 
\begin{eqnarray}
1 = d^{3d(n-1)/2}\int_{S>0}dS\,J(S)\,\exp\left(d\,\Gamma\left(S\right)\right) \,,
\end{eqnarray}
where the last integral runs over positive definite symmetric matrices $S$ and $J(S)$ is the Jacobian of the change of variables when going from the $3(n-1)d$-dimensional vector $\boldsymbol{\xi}$ to the matrix of scalar products $S$. Its expression can be found in \cite{fyodorov2007classical} and reads in the large-$d$ limit
\begin{eqnarray}
J(S) \underset{d\to\infty}{\sim} \exp\left(\frac{d}{2}\ln\det S-\frac{3d}{2}(n-1)\ln\left(\frac{d}{2\pi}\right)+\frac{3d}{2}(n-1)\right)\,.
\end{eqnarray}
The normalization condition thus becomes
\begin{equation}\label{eq:norm_cond}
\frac{1}{2}\ln\det S_{\rm sp}+\frac{3}{2}(n-1)\ln\left(2\pi\right)+\frac{3}{2}(n-1)+\,\Gamma\left(S_{\rm sp}\right) = 0
\end{equation}
with $S_{\rm sp}$ the saddle point matrix 
\begin{eqnarray}
S_{\rm sp}=\underset{S>0}{{\rm argmax}}\left(\frac{1}{2}\ln\det S+\Gamma\left(S\right)\right)\,,
\end{eqnarray}
associated to the distribution in Eq.~\eref{eq:ansatz}. As we show next, the free energy associated to any distribution satisfying the ansatz in Eq.~\eref{eq:ansatz} can be expressed, to leading order in $d$, as a function of the corresponding saddle point matrix $S_{\rm sp}$ only. This allows to recast the functional minimization of the free energy into a simpler optimization problem over the elements of $S_{\rm sp}$. 

\subsection{Computing the free energy: the ideal gas part}\label{subsec:compute_free}

For a generic distribution $\rho(\bi{r},\bxi)$, we define from Eq.~\eref{eq:free_energy} the ideal gas contribution per unit volume as
\begin{eqnarray} \label{eq:ideal_gas}
g_{\rm id}[\rho] =  \frac{1}{V}\int \rmd \bi{r} \rmd \bxi \, \rho(\bi{r},\bxi)\left[\ln \rho(\bi{r},\bxi) - 1 - j^*(\bi{r},\bxi)\right] \,.
\end{eqnarray}
Using the scaling form in Eq.~\eref{eq:ansatz}, we get
\begin{eqnarray}
\fl
g_{\rm id}[\rho] = \rho \left(\ln \rho - \mu - 1\right) + \rho d \int \rmd \bxi\, {\rm e}^{d\,\Gamma\left(\frac{S}{d}\right)} \left[\Gamma\left(\frac{S}{d}\right) + \frac{1}{2}{\rm Tr}\left(\frac{S}{d}\right) + \frac{3(n-1)}{2}\ln(2\pi) \right] \,,
\end{eqnarray}
which, after performing a saddle-point evaluation of the integral and using the normalization condition Eq.~\eref{eq:norm_cond}, reduces to 
\begin{eqnarray}
g_{\rm id}[\rho] = \rho \left(\ln \rho - \mu - 1\right) + \frac{d\rho}{2} \left({\rm Tr}S_{\rm sp} - \ln\det S_{\rm sp} - 3(n-1)\right) \,.
\end{eqnarray}

\subsection{The interacting part} 

For a generic distribution $\rho(\bi{r},\bxi)$, we define the interacting part of the free energy per unit volume as
\begin{eqnarray} \label{eq:interacting}
g_{\rm int}[\rho] =  -  \frac{1}{2V} \int \rmd \bi{r} \, \rmd \bxi \, \rmd \bi{r}' \, \rmd \bxi' \, \rho(\bi{r},\bxi) \rho(\bi{r}',\bxi') f(\bi{r}, \bxi ; \bi{r}', \bxi')\,.
\end{eqnarray}
We now restrict ourselves to distribution satisfying the ansatz in Eq.~\eref{eq:ansatz}. We introduce $r = |\bi{r}|$ and $\hat{\bi{r}} = \bi{r}/|\bi{r}|$ and perform the changes of variables $r\to\sigma(1+h/d)$, $\bxi \to \sqrt{d} \bxi$ and $\bxi' \to \sqrt{d} \bxi'$. To leading order in the dimension, we get 
\begin{eqnarray} \label{eq:interacting_bis}
g_{\rm int}[\rho] = - \frac{d\rho}{2}\hat{\varphi} \int \rmd h \, \rme^h \int \rmd \bxi \rmd \bxi' \rme^{d\hat{\Gamma}(S) + d\hat{\Gamma}(S')} I(h, \bxi, \bxi')\,,
\end{eqnarray}
with
\begin{eqnarray}
I(h, \bxi, \bxi') = \int \frac{\rmd \hat{\bi{r}}}{\Omega_d} f\left(0,\sqrt{d}\bxi;\hat{\bi{r}},h,\sqrt{d}\bxi'\right) \,.
\end{eqnarray}
Due to rotational symmetry, $I$ can be expressed as a function of $S$, $S'$ and the matrix of cross scalar products $X_{ab} = \bxi'_a \cdot \bxi_b$ for $a,b = 1, \dots, 3(n-1)$.  Because the sets of vectors $\bxi$ and $\bxi'$ are independently distributed (at the level of the large deviation function) with rotationally invariant statistics, the scalar products $R_{ab}$ typically scale as $d^{-1/2}$ and can be set to $0$ when evaluating the integrals over $\bxi$ and $\bxi'$ in Eq.~\eref{eq:interacting_bis}. We are thus left with
\begin{eqnarray} \label{eq:preliminary_fint}
g_{\rm int}[\rho] = - \frac{d\rho}{2}\hat{\varphi} \int \rmd h \, \rme^h \, I(h, S = S_{\rm sp}, S' = S_{\rm sp}, X = 0)\,.
\end{eqnarray}
We now compute $I(h, \bxi, \bxi')$. The Mayer function can be read from Eq.~\eref{eq:xi_n} as
\begin{eqnarray}
\fl f\left(0,\sqrt{d}\bxi;\hat{\bi{r}},h,\sqrt{d}\bxi'\right) = - 1 + \exp\left[-\beta\hat{U}(h) - \frac{\hat{\tau}\hat{U}'(h)}{2\sigma^2}\sum_{a=1}^{2n-2}\left|\bphi_a-\bphi'_a\right|^2 \right]\times \nonumber \\ \fl \exp\left[-\frac{\sqrt{\beta\hat{\tau}}\hat{U}'(h)}{\sigma}\sqrt{d}\sum_{a=1}^{n-1} \left(\bpsi_a - \bpsi'_a\right)\cdot\hat{\bi{r}} - \frac{\hat{\tau} \hat{U}''(h)}{2\sigma^2}d\sum_{a=1}^{2n-2}\left[\left(\bphi_a-\bphi'_a\right)\cdot\hat{\bi{r}}\right]^2\right]\,.
\end{eqnarray}
To leading order in the dimension, the integral over the unit vector $\hat{\bi{r}}$ can be expressed as a Gaussian integral. Indeed,  
\begin{eqnarray}
\fl \int \frac{\rmd \hat{\bi{r}}}{\Omega_d}\exp\left[-\frac{\sqrt{\beta\hat{\tau}}\hat{U}'(h)}{\sigma}\sqrt{d}\sum_{a=1}^{n-1} \left(\bpsi_a - \bpsi'_a\right)\cdot\hat{\bi{r}} - \frac{\hat{\tau} \hat{U}''(h)}{2\sigma^2}d\sum_{a=1}^{2n-2}\left[\left(\bphi_a-\bphi'_a\right)\cdot\hat{\bi{r}}\right]^2\right] \nonumber \\
\fl = \int \frac{\rmd \hat{\bi{r}}}{\Omega_d} \int \prod_{a=1}^{2n-2} \rmd x_a \prod_{a=1}^{2n-2} \delta\left(x_a - \sqrt{d}\left(\bphi_a-\bphi'_a\right)\cdot\hat{\bi{r}}\right) \nonumber \\ 
\fl \exp\left[-\frac{\sqrt{\beta\hat{\tau}}\hat{U}'(h)}{\sigma}\sqrt{d}\sum_{a=1}^{n-1} \left(\bpsi_a - \bpsi'_a\right)\cdot\hat{\bi{r}} - \frac{\hat{\tau} \hat{U}''(h)}{2\sigma^2}d\sum_{a=1}^{2n-2}x_a^2\right] \nonumber \\
\fl = \int \prod_{a=1}^{2n-2} \frac{\rmd x_a \rmd \hat{x}_a}{2\pi} \, \exp\left(i\sum\limits_{a=1}^{2n-2}\hat{x}_a x_a - \frac{\hat{\tau} \hat{U}''(h)}{2\sigma^2}d\sum_{a=1}^{2n-2}x_a^2\right)\nonumber \\
\fl \int \frac{\rmd \hat{\bi{r}}}{\Omega_d} \exp\left(-\sqrt{d}\hat{\bi{r}}\cdot \left[\frac{\sqrt{\beta\hat{\tau}}\hat{U}'(h)}{\sigma}\sum_{a=1}^{n-1}\left(\bpsi_a - \bpsi'_a\right) + i\sum_{a=1}^{2n-2} \hat{x}_a \left(\bphi_a - \bphi'_a\right) \right]\right) \nonumber \\
\fl = \int \prod_{a=1}^{2n-2} \frac{\rmd \hat{x}_a}{\sqrt{2\pi}} \, \exp\left(-\frac{1}{2}\sum\limits_{a=1}^{2n-2}\hat{x}^2_a \right)\nonumber \\
\fl \exp\left(\frac{1}{2}\left(\frac{\sqrt{\beta\hat{\tau}}\hat{U}'(h)}{\sigma}\sum_{a=1}^{n-1} \left(\bpsi_a - \bpsi'_a\right) + i \sqrt{\frac{\hat{\tau} \hat{U}''(h)}{\sigma^2}} \sum_{a=1}^{2n-2} \hat{x}_a \left(\bphi_a - \bphi'_a\right) \right)^2\right)\,.
\end{eqnarray}
Upon replacing $S$ and $S'$ by $S_{\rm sp}$ and setting $X = 0$, and after performing the remaining Gaussian integrals, we therefore get
\begin{eqnarray}
\fl I(h, S = S_{\rm sp}, S' = S_{\rm sp}, R = 0) = - 1 + \frac{\exp\left(-\beta\hat{U}(h) - \frac{\hat{\tau}\hat{U}'(h)}{\sigma^2}{\rm Tr}Q_{\rm sp} + \frac{\beta\hat{\tau}}{\sigma^2}\hat{U}'(h)^2 V_{n-1}^T M_{\rm sp} V_{n-1}\right)}{\sqrt{\det\left(\mathbbm{1} + \frac{2\hat{\tau} \hat{U}''(h)}{\sigma^2}Q_{\rm sp}\right)}} \times \nonumber \\ \fl\exp\left(-\frac{2\beta\hat{\tau}^2}{\sigma^4}\hat{U}''(h)\hat{U}'(h)^2 \, V_{n-1}^T R_{\rm sp} \left(\mathbbm{1} + \frac{2\hat{\tau} \hat{U}''(h)}{\sigma^2}Q_{\rm sp}\right)^{-1} R_{\rm sp}^T V_{n-1}\right) \,,
\end{eqnarray}
where $V_m$ is the $m$-dimensional vector with $V_m = 1$ for all $a = 1,\dots,m$. This expression, together with Eq.~\eref{eq:preliminary_fint}, completes our computation of the free energy.

\subsection{Replica symmetric ansatz}\label{sec:RS_ansatz}
We have shown that the free energy of any homogeneous and rotationally invariant one-body density with a large deviation form given in Eq.~\eref{eq:ansatz} can be expressed as a function of the number density $\rho$ and the corresponding saddle point matrix $S_{\rm sp}$ describing the orientational order of the internal auxiliary degrees of freedom $\{\bpsi_a, \bphi_a\}$. We have obtained (forgetting the subscript sp from now on)
\begin{eqnarray}
g_n[\rho]=\left[\rho(\ln\rho-1)-\mu\rho\right]-\frac{d\rho}{2}\tilde{g}_n(\hat{\varphi},S_{\rm})
\end{eqnarray}
with
\begin{eqnarray}
\tilde{g}_n(\hat{\varphi},S)=\ln\det S+3(n-1)-{\rm Tr}S \nonumber \\  +\hat{\varphi}\int{\rm d}h\,{\rm e}^{h}\Bigg[-1+\frac{\exp\left(-\beta \hat{U}(h)-\frac{\hat{\tau} \hat{U}'(h)}{\sigma^{2}}{\rm Tr}Q+\frac{\beta\hat{\tau}}{\sigma^{2}}\hat{U}'(h)^{2}V_{n-1}^{T}MV_{n-1}\right)}{\sqrt{\det\left(\mathbbm{1}+\frac{2\hat{\tau} \hat{U}''(h)}{\sigma^{2}}Q\right)}}\nonumber \\ \times\exp\left(-\frac{2\beta\hat{\tau}^{2}}{\sigma^{4}}\hat{U}''(h)\hat{U}'(h)^{2}\,V_{n-1}^{T}R\left(\mathbbm{1}+\frac{2\hat{\tau} \hat{U}''(h)}{\sigma^{2}}Q\right)^{-1}R^{T}V_{n-1}\right)\Bigg] \,.
\end{eqnarray}
In turn, the physical free energy at chemical potential $\mu$ results from the optimization principle
\begin{eqnarray}
g_n(\mu)=\underset{\rho}{{\rm min}}\left(\left[\rho(\ln\rho-1)-\mu\rho\right]-\frac{d\rho}{2}\underset{S}{{\rm max}} \, \tilde{g}_n(\hat{\varphi},S)\right) \,.
\end{eqnarray}
To proceed with the optimization of the free energy over the elements of $S$, we restrict our analysis to a replica symmetric ansatz
\begin{eqnarray}
M	&= m_{0}\mathbbm{1}_{n-1}+m_{1}V_{n-1}V_{n-1}^{T} \,, \nonumber \\
Q	&= q_{0}\mathbbm{1}_{2(n-1)}+q_{1}V_{2(n-1)}V_{2(n-1)}^{T} \,, \nonumber \\
R	&= rV_{n-1}V_{2(n-1)}^{T} \,.
\end{eqnarray}
To be more specific, we consider harmonic spheres $\hat{U}(h) = u_0 h^2/2 \Theta(-h)$ and define as in section \ref{sec:notransiUCNA} the parameters $c_1 = 2\hat{\tau}u_0/\sigma^2$ and $c_2 = \beta u_0$. The limit $\tilde{g}(\hat{\varphi},m_0,m_1,q_0,q_1,r) = \lim_{n \to 0}\tilde{g}_n(\hat{\varphi},m_0,m_1,q_0,q_1,r)$ then reads
\begin{eqnarray}
\fl \tilde{g}(\hat{\varphi},m_0,m_1,q_0,q_1,r)\nonumber \\ \fl	= -3\ln q_{0}-2\ln m_{0}+\ln\left((m_{0}-m_{1})(q_{0}-2q_{1})-2r^{2}\right)-3  +m_{0}+m_{1}+2(q_{0}+q_{1})
	\nonumber \\ \fl +\hat{\varphi}\left[-1+\frac{\left(1+c_{1}q_{0}\right)^{3/2}}{\sqrt{1+c_{1}(q_{0}-2q_{1})}}\int_{0}^{+\infty}{\rm d}h\,\exp\left(-\frac{h^{2}}{2}c_{2}\left[1+c_{1}(m_{0}-m_{1})-c_{1}^{2}\frac{2r^{2}}{1+c_{1}(q_{0}-2q_{1})}\right] \right.\right. \nonumber \\  -h\left[1+c_{1}(q_{0}+q_{1})\right]\Bigg)\Bigg]\,.
\end{eqnarray} 
We furthermore note that 
\begin{eqnarray}
r = 0 \Rightarrow \partial_r\tilde{g}(\hat{\varphi},S)  = 0 \,,
\end{eqnarray}
and thus restrict ourselves to the ensemble of saddle-point solutions with $r = 0$, corresponding to states where there are no correlations between the fields $\{\bpsi_a\}$ and $\{\bphi_a\}$. Lastly, we introduce the variables $m = m_0 - m_1$, $q = q_0$ and $p = q_0 - 2 q_1$, while keeping the variable $m_0$. Taking the gradient with respect to the latter then yields $m_0 = 1$ so that we are left with optimizing the function 
\begin{eqnarray}
\fl \tilde{g}(\hat{\varphi},m,p,q)	= -3\left(\ln q - q + 1\right) + \ln m - m + 1 + \ln p -p + 1 \nonumber \\ \fl \hat{\varphi}\left[-1+\frac{\left(1+c_{1}q\right)^{3/2}}{\sqrt{1+c_{1}p}}\int_{0}^{+\infty}{\rm d}h\,\exp\left(-\frac{h^{2}}{2}c_{2}\left(1+c_{1}m\right)-h\left[1+\frac{c_{1}}{2}(3q-p)\right]\right)\right] \,.
\end{eqnarray} 
In the next section, we discuss the physical implications of this optimization principle. 

\section{Phase diagram of the UCNA}\label{sec:phase_diagram}

\subsection{The saddle point equations}
We begin this section by summarizing what we have achieved so far. We started with the position-space $N$-body stationary distribution given in Eqs.~\eref{eq:UCNAstat} and \eref{eq:matrixH}. This distribution features multibody interactions which we converted into pairwise interactions by introducing auxiliary internal degrees of freedom attached to each particle. This led us to express the grand canonical partition function $\Xi[\mu]$ as the limit when $n \to 0$ of a sequence of partition functions $\Xi_n[\mu]$ given in Eq.~\eref{eq:xi_n}. At positive integers $n$, $\Xi_n[\mu]$ can be interpreted as the annealed average of the partition function of a spherical-like (with harmonic confinement) replicated spin system with short-ranged disordered pairwise interactions between the spins set by the positions of the particles. The average over the disorder is then taken by distributing those positions according to the Boltzmann weight. The collective properties are then inferred from minimizing the free energy functional over one-body distributions, which, in the limit $d \to \infty$, amounts to optimizing over the number density $\rho$ and over a spin-glass-like order parameter measuring on-site correlations between replicas of these internal vectorial degrees of freedom. By restricting ourselves to a replica symmetric ansatz (the structure of which is precisely defined in section \ref{sec:RS_ansatz}), and after taking the $n \to 0$ limit, we have shown that the free energy per unit volume (in a homogeneous phase) takes the form
\begin{eqnarray}\label{eq:min_final}
g(\mu)=\underset{\rho}{{\rm min}}\left(\left[\rho(\ln\rho-1)-\mu\rho\right]-\frac{d\rho}{2} \, \tilde{g}(\hat{\varphi},m^*, p^*, q^*)\right) \,,
\end{eqnarray}
where $m^*$, $p^*$ and $q^*$ are order parameters for the correlations of the replicated auxiliary degrees of freedom. Heuristically, in  terms of the auxiliary variables $\{\bphi_a\}$, $q^* = \left\langle \bphi_a \cdot \bphi_a \right\rangle$ measures correlations in a single replica while $q^* - p^* \propto \left\langle \bphi_a \cdot \bphi_b \right\rangle$ for $a \neq b$ measures correlations between different replicas. A state with $q^* \neq p^*$ therefore describes a spin-glass phase (in the sense of a non-vanishing Edwards-Anderson order parameter) of the (a priori unphysical) auxiliary degrees of freedom. For harmonic spheres (and using the parameters $c_1$ and $c_2$ defined in section \ref{sec:notransiUCNA}), the set of order parameters $(m^*, p^*, q^*)$ is obtained as a critical point of the function
\begin{eqnarray}\label{eq:tildeg_final}
\fl \tilde{g}(\hat{\varphi}, m, p, q)	= -3\left(\ln q - q + 1\right) + \ln m - m + 1 + \ln p -p + 1 \nonumber \\ \fl \hat{\varphi}\left[-1+\frac{\left(1+c_{1}q\right)^{3/2}}{\sqrt{1+c_{1}p}}\int_{0}^{+\infty}{\rm d}h\,\exp\left(-\frac{h^{2}}{2}c_{2}\left(1+c_{1}m\right)-h\left[1+\frac{c_{1}}{2}(3q-p)\right]\right)\right] \,.
\end{eqnarray}
At finite replica index $n$, the values of the order parameters are obtained by maximizing a finite $n$ generalization of the function $\tilde{g}$, but maxima become saddle points after taking the $n \to 0$ limit. For instance, at zero density $\hat{\varphi} = 0$, the only critical point is given by 
\begin{eqnarray}
m^* = p^* = q^* = 1 \,,
\end{eqnarray}
and it describes a paramagnetic phase. At that point, the Hessian of $\tilde{g}$ has two positive eigenvalues and a negative one. Generalizing this result to nonvanishing densities, we consider that a state is stable only if the corresponding Hessian has signature $(+, -, -)$. Interestingly, we numerically found that when several critical points coexist, the one which maximizes the function $\tilde{g}(\hat{\varphi}, m, q, p)$ is always associated with this $(+, -, -)$ signature, see \fref{fig:numerics1}(d). 
We are now equipped to derive the phase diagram of the UCNA in the infinite dimensional limit by numerically solving the saddle point equations corresponding to the function $\tilde{g}$ in Eq.~\eref{eq:tildeg_final}. Prior to that, and despite the absence of obvious symmetry between the variables $p$ and $q$ in Eq.~\eref{eq:tildeg_final}, it is interesting to note that for any density $\hat{\varphi}$, there exists a critical point (stable or unstable) of $\tilde{g}$ with $p = q$ corresponding to a paramagnetic phase. This stems for the following relation at $p = q$
\begin{eqnarray}
\left. \partial_q \tilde{g} \right|_{\hat{\varphi}, m, q, q} = - 3 \left. \partial_p \tilde{g} \right|_{\hat{\varphi}, m, q, q} \,.
\end{eqnarray} 

\subsection{Numerical solutions}

We solve numerically the saddle point equations $\partial_m \tilde{g}(\hat{\varphi}, m, p,q) = \partial_p \tilde{g}(\hat{\varphi}, m, p,q) = \partial_q \tilde{g}(\hat{\varphi}, m, p,q) = 0$. Here, we restrict ourselves to $c_2 = 0.1$ and vary the density $\hat{\varphi}$ and $c_1$ (see \ref{app:additional_figures} for the case $c_2 = 0.5$). This amounts to varying the persistence time of the original AOUPs dynamics while working at constant $\beta = D^{-1}$. \\

\begin{figure}[htbp]
    \centering
    \includegraphics[width=0.8\textwidth]{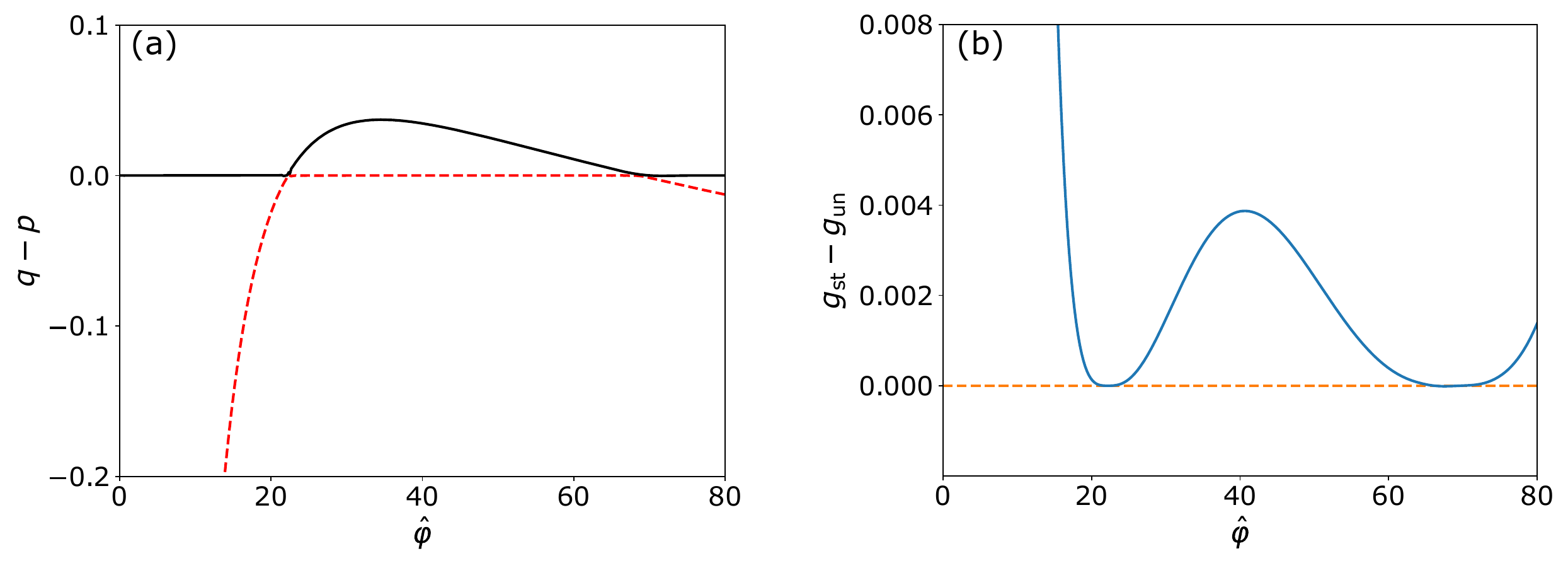}
    \caption{Numerical solution of the saddle-point equations at $c_2 = 0.1$ and $c_1 = 0.7$. At any density, there exist two branches of solutions: One corresponds to a paramagnetic phase with $q - p = 0$ and one to a spin-glass phase with $q - p \neq 0$. They intersect at a low and a high density. (a) The stable branch (continuous, black) is the paramagnetic one at low and high densities and the spin-glass one at intermediate densities. The red dashed line corresponds to the unstable branch of solution. (b) The free energy of the unstable branch $g_{\rm un}$ is always higher than the free energy of the stable one $g_{\rm st}$.}
    \label{fig:numerics1}
\end{figure}

\begin{figure}[htbp]
    \centering
    \includegraphics[width=0.8\textwidth]{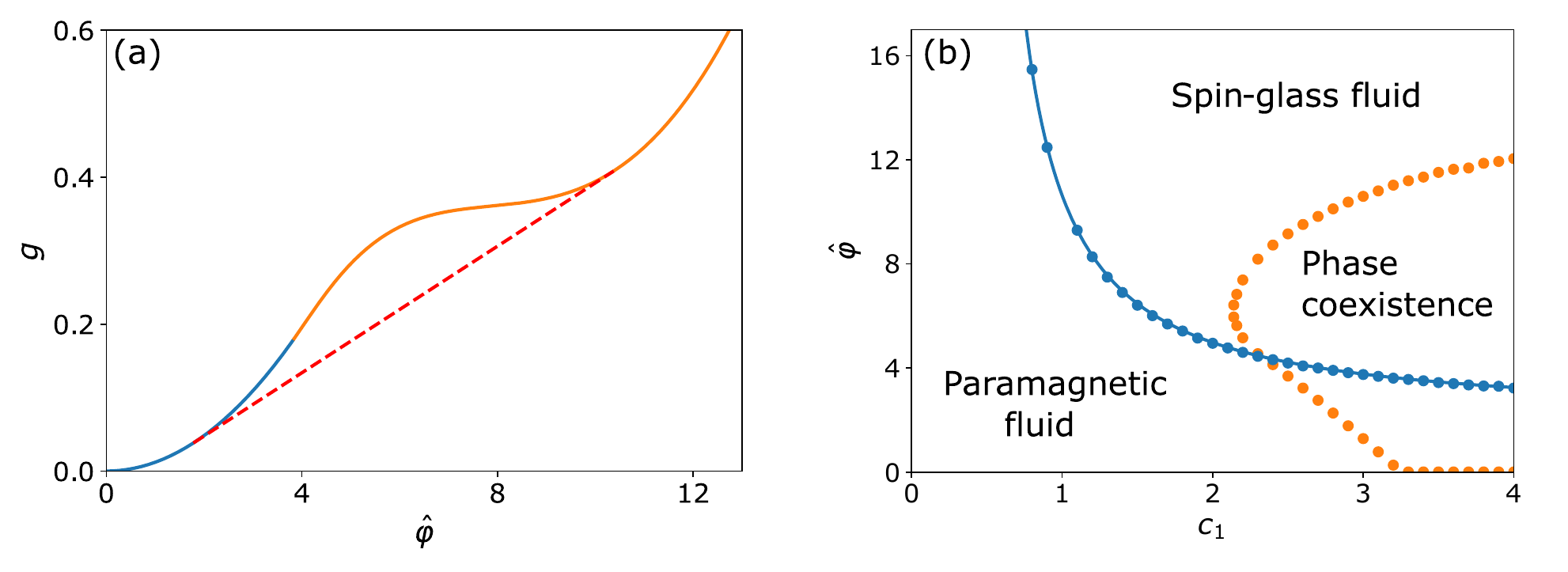}
    \caption{Phase transitions in the large-$d$ UCNA steady state. (a) Free energy as a function of the density for $c_2 = 0.1$ and $c_1 = 2.9$. The blue part of the curve corresponds to a paramagnetic state and the orange one to a spin-glass state. The dashed red line indicates the convex envelop of the free energy from which one can read the binodals. (b) Phase diagram of the UCNA. The blue line indicates the liquid-liquid continuous phase transition between the paramagnetic and the spin-glass phases (the dots were obtained from the numerical solution of the saddle point equations and the continuous line is the theoretical prediction of \ref{app:analytical_transition}). The orange dots form the binodals of the phase separation. The critical point of the MIPS-like phase transition lies well within the spin-glass phase.}
    \label{fig:phase_diagram}
\end{figure}

We find that for any finite density there exist two branches of solutions, one corresponding to a paramagnetic state with $p = q$ and one to a spin glass state with $p \neq q$. At small persistence times, for $c_1 \leq c_1^{\rm sg} \simeq 0.6 $, the spin glass solution is unstable at any density $\hat{\varphi}$ and the paramagnetic one is always stable. At larger persistence times $c_1 \geq c_1^{\rm sg}$, the two branches of solution intersect at a low and a high density. In the intermediate density region, the paramagnetic branch loses stability and the spin glass state becomes the stable one, see \fref{fig:numerics1}. Note that to leading order in $d$, the usual ideal gas term is negligible and, up to the linear contribution $\mu\rho$, the free energy can be expressed as a function of the density as
\begin{eqnarray}
g(\hat{\varphi}) \simeq -\frac{d^2\hat{\varphi}}{2\mathcal{V}_d(\sigma)} \, \tilde{g}(\hat{\varphi},m^*, p^*, q^*) \,,
\end{eqnarray}
so that numerical plots of the free energy correspond to plotting $-(\hat{\varphi}/2) \tilde{g}(\hat{\varphi},m^*, p^*, q^*)$. We will come back to that point in subsection \ref{sec:real}. In the $(c_1, \hat{\varphi})$ plane, this defines a transition line between the paramagnetic and the spin glass phases, for which the analytical prediction obtained in \ref{app:analytical_transition} matches the results of numerical optimization, see \fref{fig:phase_diagram}(b). As we show in subsection \ref{sec:real}, this transition, which is continuous in the order parameters $m^*$, $p^*$ and $q^*$, manifests itself as a continuous liquid-liquid phase transition in position space with a discontinuity in the derivative  with respect to the mean density of the relative number fluctuations ($\left\langle \Delta N^2 \right\rangle/\left\langle N \right\rangle$ with $N$ the number of particles in a large volume $V$ and $\Delta N = N - \left\langle N \right\rangle$). That the Edwards-Anderson order parameter vanishes above a certain density suggests that there, the disorder induced by density fluctuations is not strong enough to support a spin-glass phase. \\

At even larger persistence times, $c_1 \geq c_1^{\rm cr} \simeq 2.1$, the free energy becomes a non-convex function of the density, see \fref{fig:phase_diagram}(a). This non-convexity signals the emergence of a MIPS-like phase separation. The corresponding critical point lies well within the spin-glass phase and therefore belongs to the (mean-field) Ising universality class. At moderate values of  $c_1$, the system can phase separate between two spin-glass phases of different densities. As the persistence time is increased, the lower binodal crosses the phase separation line between the spin-glass and the paramagnetic phases and the system phase separates between a dense spin-glass and a dilute paramagnet. At even higher persistence times, the low-density binodal saturates at $\hat{\varphi} = 0$, corresponding to a very dilute phase with less than $O(d)$ neighbors per particle. The full phase diagram is reported in \fref{fig:phase_diagram}(b). For the sake of visual clarity, we restricted the plot to moderate density values, but an extended version of the phase diagram can be found in \ref{app:additional_figures} where we also present the phase diagram obtained for $c_2 = 0.5$. It shows a similar phenomenology with a shift towards larger persistence times of the MIPS critical point and of the phase boundary between the paramagnet and the spin-glass. 

\subsection{Real space signatures of the paramagnet to spin glass transition}\label{sec:real}
\begin{figure}[htbp]
    \centering
    \includegraphics[width=0.6\textwidth]{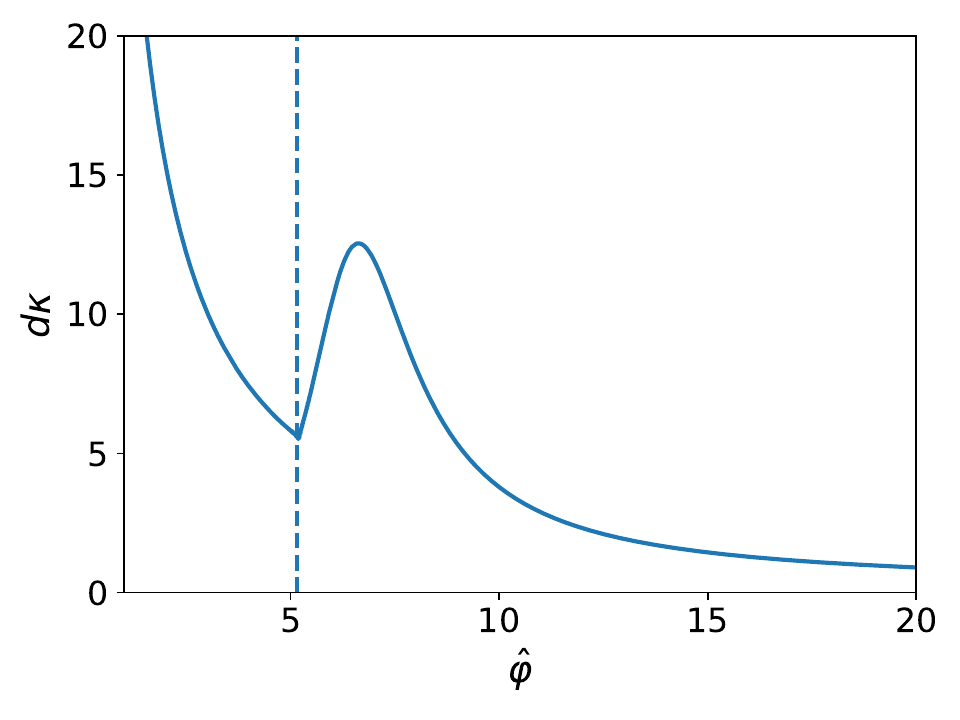}
    \caption{Number fluctuations $\kappa$ as a function of the reduced density $\hat{\varphi}$ for $c_1 = 1.9$ and $c_2 = 0.1$. They are continuous with discontinuous derivative upon crossing the paramagnet-to-spin-glass transition marked by the dashed vertical line. Note that for the ideal gas, $\kappa = 1$ so that $d\kappa$ diverges in the limit of infinite dimension when $\hat{\varphi} \to 0$. The secondary peak signals the proximity of the MIPS critical point.}
    \label{fig:kappa}
\end{figure}

In order to assess the effect of the paramagnet-to-spin-glass transition on real space physics, we now discuss the behavior of number fluctuations. In the grand canonical ensemble, fluctuations of the number $N$ of particles in a large volume obey
\begin{eqnarray}
\kappa = \frac{\left\langle \left(\Delta N\right)^2 \right\rangle}{\left\langle N \right\rangle} = \rho^{-1}\partial_\mu \rho = \hat{\varphi}^{-1}\partial_\mu \hat{\varphi}\,,
\end{eqnarray} 
where we assumed that the system is homogeneous with mean density $\rho$ and where $\Delta N = N - \left\langle N \right\rangle$. At a given chemical potential $\mu$, the density is set by the minimization principle in Eq.~\eref{eq:min_final}. When written in terms of the rescaled chemical potential $\hat{\mu} = (\mu - \ln(d/\mathcal{V}_d(\sigma)))/d$, the latter leads to  
\begin{eqnarray}
\hat{\varphi}(\hat{\mu}) = \underset{\hat{\varphi}}{{\rm argmin}}\left( \hat{\varphi}\ln\hat{\varphi} - \hat{\varphi} - d\hat{\mu}\hat{\varphi} - \frac{d\hat{\varphi}}{2}\tilde{g}(\hat{\varphi},m^*, p^*, q^*)\right)\,,
\end{eqnarray}
where $m^*$, $p^*$ and $q^*$ are seen as functions of $\hat{\varphi}$. To leading order as $d\to\infty$, this thus allows us to characterize number fluctuations from
\begin{eqnarray}
\partial_\mu \hat{\varphi} = \frac{1}{d}\left[\frac{{\rm d}^2}{{\rm d}\hat{\varphi}^2}\left(-\frac{\hat{\varphi}}{2}\tilde{g}(\hat{\varphi},m^*, p^*, q^*)\right)\right]^{-1} \,.
\end{eqnarray}
As expected, number fluctuations therefore diverge at the onset of the MIPS-like phase separation where the second derivative of the function $\hat{\varphi}\tilde{g}(\hat{\varphi},m^*, p^*, q^*)$ vanishes. On the contrary, number fluctuations remain finite at the paramagnet-to-spin-glass transition. They are also continuous upon crossing the transition line, even though the derivatives of $m^*$, $p^*$ and $q^*$ with respect to the density are not. In fact, the saddle point equations satisfied by $m^*$, $p^*$ and $q^*$ impose at the transition
\begin{eqnarray}
\frac{{\rm d}}{{\rm d}\hat{\varphi}}\left(-\frac{\hat{\varphi}}{2}\tilde{g}\right) = \partial_{\hat{\varphi}}\left(-\frac{\hat{\varphi}}{2}\tilde{g}\right)\,,
\end{eqnarray}
and
\begin{eqnarray}
\frac{{\rm d}^2}{{\rm d}\hat{\varphi}^2}\left(-\frac{\hat{\varphi}}{2}\tilde{g}\right) = \partial^2_{\hat{\varphi}}\left(-\frac{\hat{\varphi}}{2}\tilde{g}\right) + \frac{{\rm d}m_\mu}{{\rm d}\hat{\varphi}}\partial_{m_\mu} \partial_{\hat{\varphi}}\left(-\frac{\hat{\varphi}}{2}\tilde{g}\right) \,,
\end{eqnarray}
where the index $\mu$ runs from $1$ to $3$ and where we used the reduced notation $m_1 = m$, $m_2 = p$ and $m_3 = q$. We further note that the saddle point equation $\partial_{m_\mu}\tilde{g} = 0$ leads to
\begin{eqnarray} \label{eq:Hessian_equation}
H_{\mu\nu}\frac{{\rm d}m_\nu}{{\rm d}\hat{\varphi}} = - \partial_{m_\mu} \partial_{\hat{\varphi}}\left(-\frac{\hat{\varphi}}{2}\tilde{g}\right) \,,
\end{eqnarray}
with the Hessian matrix
\begin{eqnarray}
H_{\mu\nu} = \partial_{m_\mu} \partial_{m_\nu}\left(-\frac{\hat{\varphi}}{2}\tilde{g}\right) \,.
\end{eqnarray}
The transition corresponds to points where the two branches of solutions intersect and exchange stability, and thus where the Hessian $H_{\mu\nu}$ has a zero mode. As shown by Eq.~\eref{eq:Hessian_equation}, the discontinuity in ${\rm d}m_\mu/{\rm d}\hat{\varphi}$ between the two branches of solution must be along that zero mode of the Hessian matrix (the right-hand-side is continuous at the transition). By taking the dot product between Eq.~\eref{eq:Hessian_equation} and ${\rm d}m_\mu/{\rm d}\hat{\varphi}$, we obtain that number fluctuations are indeed continuous at the transition, despite the discontinuity of ${\rm d}m_\mu/{\rm d}\hat{\varphi}$. However, the derivative of the number fluctuations with respect to the density is not, see \fref{fig:kappa}. In real space, the paramagnet to spin-glass transition is therefore akin to a continuous liquid-liquid phase transition. 

\section{Conclusion}
Within the UCNA, it is possible to determine the stationary  distribution of a system of self-propelled particles interacting by means of pairwise forces. The resulting distribution displays multibody interactions. By working in infinite space dimension we are able to extract several interesting physical features. The first one is an analytical confirmation of the numerical findings of \cite{turci2021phase}: the effective pair potential of the stationary distribution is unable to account on its own for the existence of the motility-induced phase separation (in spite of its well-known appealing features such as effective attraction). Second, our phase diagram is significantly more complex than expected. Indeed, we identify the existence of two homogeneous fluid phases. Our derivation led us to introduce replicated auxiliary internal vectorial degrees of freedom attached to each particle. At a given set of particle positions, the replicas are independent and the auxiliary degrees of freedom between different particles are coupled in a disordered way, the disorder being set by the positions of the particles themselves. The two phases then correspond to a paramagnetic or spin glass order for these auxiliary (and a priori unphysical) degrees of freedom, in the sense of a zero or nonzero Edwards-Wilkinson order parameter after averaging over the particle positions. At low persistence time, we find a single liquid paramagnetic phase. At higher persistence time, there exists a continuous transition between the paramagnetic and the spin-glass phase. In positional space, this liquid-liquid phase transition is accompanied by a discontinuity of the derivative of the compressibility. At even higher persistence time, we recover a MIPS-like first order transition.

The existence of two thermodynamically distinct liquid phases was never mentioned in previous works on self-propelled particles. The extent to which this is an artifact of the UCNA or of the infinite-dimensional limit or whether this is the signature of a real physical phenomenon is an open question.  It is tempting to speculate that the auxiliary vectorial variables we introduced are remnants of the orientation of the self-propulsion force (or of the velocities) and to think about the following scenario. Take several realizations of the steady-state with fixed positional degrees of freedom, then the two phases would correspond respectively to a state where each particle's orientations (in each copy) randomly fluctuates or to a state where it freezes in a direction selected by density fluctuations.

\ack{The authors warmly thank Julien Tailleur for countless discussions on the nature of effective interactions in active matter. FvW acknowledges the financial support of the Agence Nationale de la Recherche grant THEMA No 20-CE30-0031-01.}

\newpage 

\appendix

\section{Additional figures} \label{app:additional_figures}

\begin{figure}[h]
    \centering
    \includegraphics[width=0.8\textwidth]{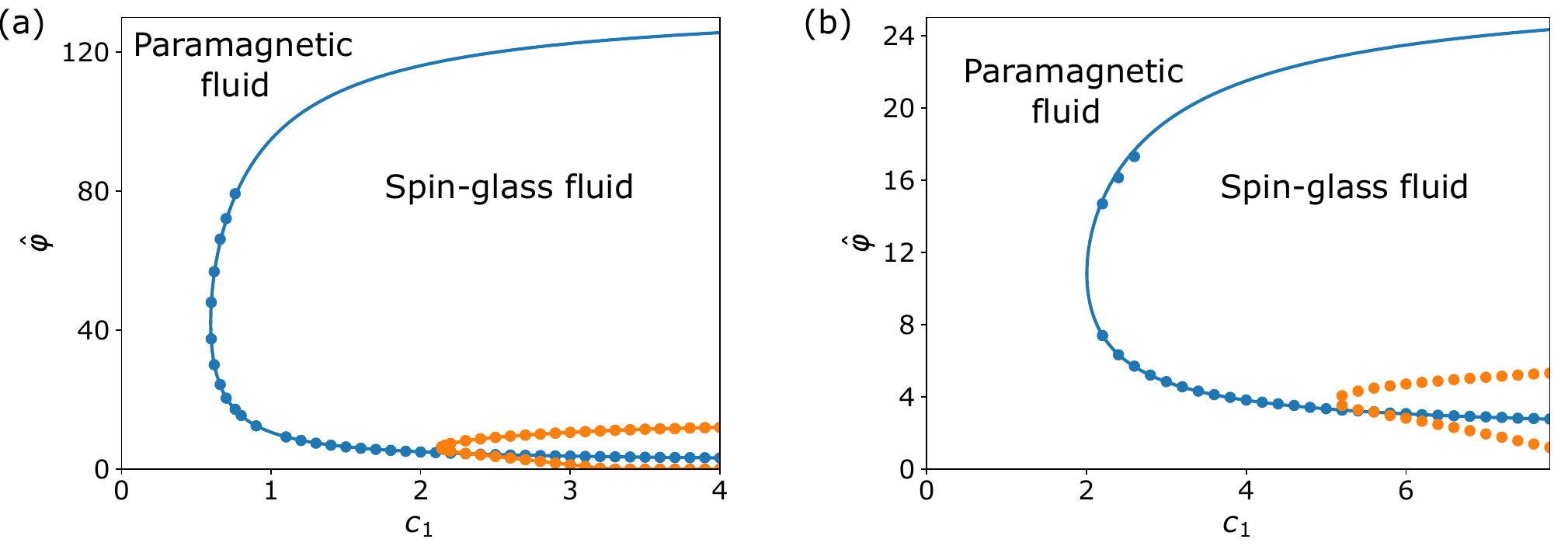}
    \caption{Phase diagrams for $c_2 = 0.1$ (a) showing higher density values than in the main text and for $c_2 = 0.5$ (b). The blue line indicates the boundary between the paramagnetic and the spin-glass phases (the dots were obtained from the numerical solution of the saddle point equations up to moderate density values and the continuous line is the theoretical prediction of \ref{app:analytical_transition}). The orange dots enclose the domain of phase coexistence. As $c_2$ increases, there is a shift towards larger persistence times of the MIPS critical point and of the phase boundary between the paramagnet and the spin-glass.}
    \label{fig:additional_phase_diagram}
\end{figure}

\section{Phase boundary of the paramagnet to spin-glass transition} \label{app:analytical_transition}
The order parameters $(m^*, p^*, q^*)$ are obtained as solutions of the saddle-point equations $\partial_m \tilde{g}(\hat{\varphi}, m, p,q) = \partial_p \tilde{g}(\hat{\varphi}, m, p,q) = \partial_q \tilde{g}(\hat{\varphi}, m, p,q) = 0$. We start by characterizing the paramagnetic branch, that is constraining the optimization to $p=q$. As noted in the main text, 
\begin{eqnarray}
\left. \partial_q \tilde{g} \right|_{\hat{\varphi}, m, q, q} = - 3 \left. \partial_p \tilde{g} \right|_{\hat{\varphi}, m, q, q} \,,
\end{eqnarray} 
which therefore leaves us with a set of two equations. Furthermore, we note that
\begin{eqnarray}
\frac{1 + c_1 m}{1 + c_1 q} \left. \partial_m \tilde{g} \right|_{\hat{\varphi}, m, q, q} -  \left. \partial_p \tilde{g} \right|_{\hat{\varphi}, m, q, q} = -(m-q)\frac{1+c_1 m q}{m q (1 + c_1 q)}\,.
\end{eqnarray}
Therefore, along the paramagnetic branch, we also have $m^* = p^* = q^*$. The latter is thus uniquely defined by the solution to the equation $\partial_m \tilde{g}(\hat{\varphi}, m, m, m) = 0$. To study the spin-glass branch, we introduce the function
\begin{eqnarray}\fl
H(\hat{\varphi}, m, p, q) = \frac{4 p q (1 + c_1 p) \sqrt{c_2 + c_1 c_2 m}}{3(p-q)}\left(\left. \partial_q \tilde{g} \right|_{\hat{\varphi}, m, p, q} + 3 \left. \partial_p \tilde{g} \right|_{\hat{\varphi}, m, p, q}\right) \,.
\end{eqnarray}
The spin-glass branch is obtained as the solution of 
\begin{eqnarray}
\left. \partial_m \tilde{g} \right|_{\hat{\varphi}, m, p, q} = \left. \partial_p \tilde{g} \right|_{\hat{\varphi}, m, p, q} = H(\hat{\varphi}, m, p, q) = 0 \,.
\end{eqnarray} 
We are now interested in the set of points where the spin-glass branch intersects the paramagnetic one. This amounts to solve for
\begin{eqnarray}
\left. \partial_m \tilde{g} \right|_{\hat{\varphi}, m, m, m} = H(\hat{\varphi}, m, m, m) = 0 \,,
\end{eqnarray}
which, at given $c_1$ and $c_2$, yield the values of $m$ and $\hat{\varphi}$ at which the paramagnet to spin glass transition takes place. We remark that 
\begin{eqnarray}
& 2 c_1 c_2 m^2 \left. \partial_m \tilde{g} \right|_{\hat{\varphi}, m, m, m} + \frac{1 + c_2 + c_1 m}{2\sqrt{c2 + c_2 c_1 m}}H(\hat{\varphi}, m, m, m) \nonumber \\ & = -2 c_2 \left(1 + c_1 m^2\right) - c_1 m (c_1 m (2 - \hat{\varphi}) + 4) -2 \,.
\end{eqnarray}
For $c_1(\hat{\varphi}-2)-2c_2 > 0$, the above second order polynomial admits a single positive root $\tilde{m}(c_1, c_2, \hat{\varphi})$. The phase boundary between the paramagnet and the spin-glass can therefore be found as the zeroes of the function $H(\hat{\varphi}, \tilde{m}(c_1, c_2, \hat{\varphi}), \tilde{m}(c_1, c_2, \hat{\varphi}), \tilde{m}(c_1, c_2, \hat{\varphi}))$ which can be found numerically. 

\section*{References}
\bibliographystyle{iopart-num}
\bibliography{biblio}

\end{document}